\documentclass[fleqn]{annalen}
\usepackage{graphics}
\begin{document}
\newcommand{\volume}{8}              
\newcommand{\xyear}{1999}            
\newcommand{\issue}{5}               
\newcommand{\recdate}{29 July 1999}  
\newcommand{\revdate}{dd.mm.yyyy}    
\newcommand{\revnum}{0}              
\newcommand{\accdate}{dd.mm.yyyy}    
\newcommand{\coeditor}{ue}           
\newcommand{\firstpage}{1}         
\newcommand{\lastpage}{???}          
\def\xc{x_{\rm c}}
\setcounter{page}{\firstpage}        
\newcommand{\keywords}{finite size scaling, Anderson transition, universality} 
\newcommand{\PACS}{71.30.+h, 71.23.-k, 72.15.Rn}
\newcommand{\shorttitle}{T. Ohtsuki {\it et al.},
Review of recent progress on numerical studies of Anderson transition}
\title{Review of recent progress on numerical studies of the Anderson transition}
\author{Tomi Ohtsuki$^{1}$, Keith Slevin$^{2}$ and Tohru Kawarabayashi$^{3}$} 
\newcommand{\address}
  {$^{1}$Department of Physics, Sophia University,
Kioicho 7-1, Tokyo 102-8554, JAPAN\\
  $^{2}$ Department of Physics, Osaka University,
Machikaneyama, Toyonaka 560-8531, JAPAN\\
  $^{3}$ Department of Physics, Toho University,
Miyama 2-2-1, Funabashi 274-8510, JAPAN}
\newcommand{\email}{\tt ohtsuki@sophia.ac.jp} 
\maketitle
\begin{abstract}
A review of recent progress in numerical studies of
the Anderson transition
in three dimensional systems is presented.
From high precision calculations the critical exponent $\nu$ for
the divergence of the localization length is estimated to be
$\nu=1.57\pm 0.02$ for the orthogonal universality class,
which is clearly distinguished from $\nu=1.43\pm 0.03$ for the
unitary universality class.
The boundary condition dependences of some quantities at the Anderson
transition are also discussed.
\end{abstract}

\section{Introduction}
It is more than four decades since Anderson pointed out the
existence of a disorder induced metal-insulator transition
(the Anderson transition) \cite{anderson}.
After the discovery of the scaling behavior of coductance
\cite{wegner,gang}, it was realised that the transition is a
second order phase transition.
This naturally leads us to expect a degree of universality,
i.e., the critical behavior does not depend on
the details of the model but depends only on the basic symmetries
of the system such as time reversal symmetry (TRS).

Numerical calculations have played a very important role in the
investigation of random systems, for example percolating systems
or spin glasses, 
where a quantitative understanding cannot be obtained
with analytic methods.
This is also true of the Anderson transition.
In this paper, we review recent progress in 
numerical studies of the Anderson transition.
Special emphasis is put on the universality of the exponent $\nu$ for
the divergence of the typical length scales.
We discuss the universal behavior of the statistics
of energy levels and conductance at the transition.
We finally comment on the boundary condition dependence
of certain quantities such as level statistics,
the conductance distribution function, and localization length.

\section{Critical exponents}
The Anderson transition is characterized by a vanishing conductivity,
a diverging dielectric constant, and a divergence of the length scales
as we change the parameter $x$ such as Fermi energy (electron density),
impurity concentration, and pressure.
Near the critical value $\xc$,
the vanishing of conductivity $\sigma$ is described by the
critical exponent $s$
\begin{equation}
  \sigma \sim (x-\xc)^s,
\label{eq_conexp}
\end{equation}
and the divergence of the dielectric constant $\varepsilon$
\begin{equation}
  \varepsilon \sim \frac{1}{(\xc-x)^{s'}}.
\label{eq_dieexp}
\end{equation}
Here we suppose that $x> (<)\xc$ is the metallic (insulating) regime.
In the metallic regime, the characteristic length scale $\xi$ is the
correlation length, while in the insulating regime, it is the
localization length.  Both diverge with the same exponent $\nu$,
\begin{equation}
  \xi \sim \frac{1}{|x-\xc|^{\nu}}.
\label{eq_locexp}
\end{equation}

From the scaling theory \cite{wegner,gang,kawabata_scal},
the conductivity exponent $s$
and the length scale exponent $\nu$ are related
\begin{equation}
 s=(d-2)\nu
\label{eq_wegner}
\end{equation}
where $d$ is the dimensionality of the system.
This is called Wegner's scaling law.
Numerical finite size scaling studies allow
$\nu$ to be accurately determined, while in the experiments $s$ is
measured.

The most accurate way of estimating the exponent $\nu$ is
the transfer matrix method used in conjunction with finite size 
scaling\cite{MK,KM}.
For three dimensional (3D) systems, we
consider a very long bar with cross section $L\times L$.
From the exponential decay length $\xi_L$ along the bar
we define the dimensionless quantity
\begin{equation}
\Lambda_L=\frac{\xi_L}{L} ,
\end{equation}
and assume a single parameter scaling form,
\begin{equation}
\Lambda_L=f(\xi/L)=F(\delta x L^{1/\nu}) .
\label{eq_sps}
\end{equation}
where $\delta x=(x-\xc)/\xc$ is the distance from the critical point.
Assuming that $\Lambda_L$ is an analytic function of $x$ for finite
$L$, we expand (\ref{eq_sps}) as
\begin{equation}
\Lambda_L=\Lambda_{\rm c} +a_1 L^{1/\nu}(x-\xc)+a_2 [L^{1/\nu}(x-\xc)]^2+\cdots .
\label{eq_expansion}
\end{equation}
Fitting the numerical data to (\ref{eq_expansion}), we estimate
$\Lambda_{\rm c},\nu,\xc$ and the expansion coefficients $a_i$.

In some cases, such as quantum percolation where the
lattice structure is disordered, the transfer matrix method
is difficult to apply and we have to use other quantities
which obey the single parameter scaling law (\ref{eq_sps}).
The analysis of energy level statistics is free from
the lattice structure problem.
For example, when we diagonalize a $L\times L\times L$ cube
and define $\Lambda_L$ by integrating
the nearest neighbor level spacing $P(s)$ up to some point $s_0$
\begin{equation}
\Lambda_L=\int_0^{s_0}P(s){\rm d}s ,
\end{equation}
then the same scaling law (\ref{eq_sps}) is valid.
In the insulating regime, in the limit of large $L$, $\Lambda_L$ approaches
$\int_0^{s_0} P_{\rm Poisson}(s){\rm d}s$
where $P_{\rm Poisson}(s)=
{\rm e}^{-s}$ is the
Poisson distribution.
On the metallic side, it approaches
$\int_0^{s_0} P_{\rm WD}(s){\rm d}s$ where $P_{\rm WD}(s)$
is the nearest neighbor spacing
for Wigner-Dyson statistics.
By making full use of scaling behavior, the exponent
$\nu$ has been estimated for various models, and the universality
of the transition confirmed with high precision.

\subsection{3D Anderson model}
The Anderson model has the Hamiltonian
\begin{equation}
 H = \sum_{<i,j>} V \exp ({\rm i}\theta_{i,j}) C_i^{\dagger}C_j +
     \sum_i W_i C_i^{\dagger}C_i ,
\end{equation}
where $C_i^{\dagger}(C_i)$ denotes a creation operator of an 
electron at the site $i$ on the 3D cubic lattice and $W_i$ denote 
the random scalar potential at the site $i$.
If the transition is universal,
even when the distribution function of the random potential is
changed, the critical exponent should be invariant.
To demonstrate this, we set all the phases $\theta_{i,j}$ to be zero
so that the time reversal symmetry exists(orthogonal universality
class), and consider
three different types of random potential:
the box distribution
\[
p(W_i)=\cases{
1/W & ( $|W_i| \leq W/2$ ) , \cr
  0 & ( otherwise ) ,
}
\]
the Gaussian distribution
\[
p(W_i) = \frac{1}{\sqrt{2 \pi \sigma^2}} \exp\left( - \frac{W_i^2}{2 \sigma^2}
\right) ,
\]
with $\sigma^2=W^2/12$,
and the Lloyd model in which $W_i$ has a Lorentz distribution
\[
p(W_i) = \frac{W}{\pi \left( W^2 + W_i^2 \right)} .
\]
For this distribution all moments higher than the mean
are divergent and the
parameter $W$ is proportional to the
full width at half maximum
of the distribution.

We then analyze the case when time reversal symmetry is broken
by magnetic fields.
If the transition is universal, the critical exponent should be independent
of how we break the time reversal symmetry.
We apply uniform magnetic fields and random magnetic fields.
In the former case, the phases of the hopping are chosen so that
the flux per unit square in the $xy$-plane is 1/3 and 1/4 of the
flux quantum, while in the latter
the phases of the hopping are random.
In Table \ref{table},
we summarize the results of the critical exponents for various cases of the
Anderson model.
From this table, we see that $\nu=1.57\pm 0.02$ for the
system with TRS.
When the TRS is broken (unitary universality class) either
by uniform or random magnetic fields, $\nu$ becomes
$1.43\pm 0.04$, clearly distinguished from that in the
presence of TRS.

\begin{table}[hc]
\begin{tabular}{|l|l|l|l|l|} \hline
     & $W_c$          & $\Lambda_{\rm c}$    &  $\nu$      & Ref.\\ \hline
OB  & 16.54(52,56) & 0.576(73,78) & 1.57(55,60) & \cite{SO2} \\
OG & 21.29(27,32) & 0.576(73,78) & 1.58(55,61) & \cite{SO2} \\
OL & 4.27(25,28) & 0.579(76,88) & 1.58(46,65)  & \cite{SO2} \\
flux 1/3 & $18.316\pm .016$  &$0.5683\pm .0013$ & $1.43\pm .04$ & \cite{SO}   \\
flux 1/4 &  $18.376\pm .017$ &$0.5662\pm .0016$ & $1.43\pm .06$& \cite{SO}   \\
random flux &$18.80 \pm .04$&$0.558 \pm .003$& $1.45 \pm .09$& \cite{KKO} \\ \hline
\end{tabular}  
\caption{The best fit estimates of the critical disorder $W_c$,
$\Lambda_{\rm c}$
and the critical
exponent and their $95\%$ confidence intervals.
OB, OG, OL are for the system invariant under the operation of the
time reversal with site potential distribution function
box, Gaussian and Lorentz, respectively. 
In ref.\cite{SO2}, corrections to scaling are taken into account.}
\label{table}
\end{table}

\subsection{Other models}
If the Anderson transition is truely universal, the exponents should be the
same, irrespective of whether we use the Anderson tight binding
model or not.
One example is the
stack of quantum Hall layers,
where electrons are allowed to hop from one layer to another \cite{OOK,CD}.
The critical exponent $\nu$ is estimated to be $1.45\pm 0.15$,
\footnote{We usually use one standard deviation for error bar except 
in Table \ref{table}.}
consistent with that in the case of Anderson model with broken time
reversal symmetry.

Another important example is that of quantum percolation
where the Hamiltonian is
\begin{equation}
H=\sum_{\langle ij \rangle}(t_{ij}C_i^{\dagger}C_j+{\rm h.c}) ,
\end{equation}
with the transfer 
integral
\begin{equation}
t_{ij}=\cases{
        V\exp({\rm i}\theta_{ij})   & ( for connected bond ), \cr
        0 & ( for disconnected bond ) ,
}
\end{equation}
Here we consider the bond percolation problem.
Bonds are
randomly connected with probabilities $p$. $\theta_{ij}$ is the Peierls phase
due to magnetic fields.
The underlying lattice is a 
three-dimensional cube of length $L$ with periodic boundary conditions
imposed.
As we increase $p$, we reach a geometrical percolation threshold
$p_{\rm c}$ and an infinite cluster is formed,
however the wave functions are still localized and the system
remains an insulator.
If we further increase the probability $p$ and
reach $p_{\rm q}$, the wave functions
become delocalized,
and for $p>p_{\rm q}$ current can flow through the system.
This is the quantum percolation.
Near the quantum percolation threshold $p_{\rm q}$,
the length scale diverges as $\xi\sim |p-p_{\rm q}|^{-\nu}$.

As mentioned above, this model is difficult to study with the transfer
matrix method, so the energy level statistics are used instead \cite{BA}.
The estimate of $\nu$ is consistent with that of the Anderson model,
$1.45\pm .11$ with TRS ($\theta_{ij}=0$), and $1.25\pm .08$ in the
absence of TRS where $\theta_{ij}$ is randomly distributed between
$-\pi$ and $\pi$.
The critical exponent indicates that
quantum percolation may be in the same universality class as the
Anderson transition \cite{kaneko}.

\section{Scale invariance and boundary condition dependence at the transition}
\subsection{Fractal dimension}

We now turn our attention to the properties
just at the Anderson transition.
It is well known that at the transition,
the wave function shows multifractal structure
\cite{aoki,SE,schreiber,OKO,SG2,PJ,BHS} which
leads to the scale invariant behavior of conductance
distributions\cite{shapiro,markos,SO,soukoulis,SO3,markos2}
and the energy level
statistics \cite{SSSLS,OO,HS,HS2,evangelou,ZK,BSZK,SZ,KOSO}.

The direct way to investigate the wave functions is to
diagonalize the Hamiltonian.
This, however, is strongly constrained by the
limited memory and CPU time.
Instead, we calculate here the time evolution of wave
packets to extract the information of fractal dimension\cite{WK}.
We first prepare the initial wave packet $|0\rangle$
close to the critical point by diagonalizing a small cluster located
at the center of the system.
The time evolution of the state at time $t$ is then
obtained by
\[
|t+\Delta t\rangle =U(\Delta t)
|t\rangle
\]
where $U(\Delta t)$
is the time evolution operator.
We approximate $U(\Delta t)$ by a product of exponential
operators
\begin{equation}
U(\Delta t) ={\rm e}^{-{\rm i}H\Delta t/\hbar}
= U_2(p\Delta t) U_2((1-2p)\Delta t) U_2(p\Delta t) 
 +{\rm O}(\Delta t^5)
\end{equation}
with $p=(2-2^{1/3})^{-1}$ and
\begin{equation}
U_2(\Delta t) \equiv {\rm e}^{-{\rm i}H_1\Delta t/2\hbar}\cdots
{\rm e}^{-{\rm i}H_{q-1}\Delta t/2\hbar}
{\rm e}^{-{\rm i}H_q\Delta t/\hbar}{\rm e}^{-{\rm i}H_{q-1}\Delta t/2\hbar}
\cdots{\rm e}^{-{\rm i}H_1\Delta t/2\hbar} , 
\end{equation}
where $H_1,\cdots,H_q$ is a decomposition of the
original Hamiltonian $H=\sum_i H_i$ in which each $H_i$ is simple enough
to be diagonalized analytically \cite{suzuki,suzuki2,suzuki3,suzuki4}.

The square displacement of a wave packets is defined by
\begin{equation}
r^2(t)=\langle t| r^2|t \rangle .
\end{equation}
In metallic phase, $r^2(t)$ is proportional to $D t$ where
$D$ is the diffusion coefficient.
In the insulating phase, it saturates to the square of localization
length, $\xi^2$.
At the critical point, the anomalous diffusion\cite{shapiro2,OK}
\begin{equation}
r^2(t)\sim t^{2/d}=t^{2/3}
\label{eq_anomdif}
\end{equation}
is expected.

The fractal dimension $D_2$ is estimated from the
autocorrelation function
\begin{equation}
C(t) =\frac{1}{t}\int_0^t {\rm d}t'
|\langle t'|0\rangle|^2 .
\end{equation}
Since $C(t)$ represents the inverse of the volume of the
wave packet at time $t$,
\[
C(t)\sim r^{-D_2} ,
\]
from (\ref{eq_anomdif}) we obtain
\begin{equation}
C(t)\sim t^{-D_2/d}.
\label{eq_autocorr}
\end{equation}
Eq. (\ref{eq_autocorr}) can be derived more rigorously in
Fourier space \cite{BHS}.

\begin{figure}
\centerline{\resizebox{6cm}{6cm}{\includegraphics{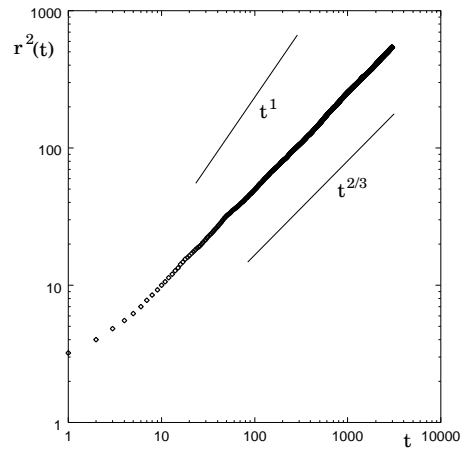}}}
 \caption{$r^2(t)$ vs. $t$ at the critical point.
Anomalous diffusion of $r^2\sim t^{2/3}$ is observed.}
 \label{xi_eps_fig}
\end{figure}
\begin{figure}
\centerline{\resizebox{6cm}{6cm}{\includegraphics{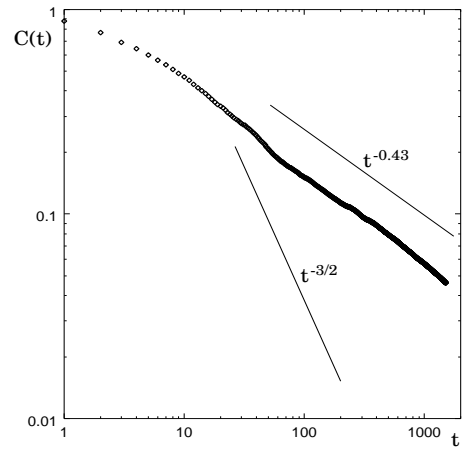}}}
 \caption{Autocorrelation function $C(t)$ as a function of time.
Due to the fractal structure of the wave function, the decay of $C(t)$ is slow,
$\sim t^{-0.43}$, compared to $t^{3/2}$ expected for normal diffusion.}
 \label{ret_eps_fig}
\end{figure}

In Figs. \ref{xi_eps_fig} and \ref{ret_eps_fig},
we show the results of $r^2(t)$ and $C(t)$
at the transition in the
presence of a uniform magnetic field.
The strength of the field is 1/4 magnetic flux per
unit square lattice in the $xy$-plane.
The Anderson transition takes place at $W_{\rm c}=18.4V$.
By diagonalizing a small cluster of $7\times 7\times 7$ located at the
the center of the system,
we follow the time evolution of wave
packets in $101\times 101\times 101$ systems.
The geometric average of $C(t)$ over 10
potential configurations are performed.
By fitting the data for $t>40 \hbar/V$,
the fractal dimensionality $D_2$ is estimated to be
\[
D_2 = 1.3\pm 0.2
\]
considerably smaller than the space dimension $d=3$.
The above value is consistent with the estimate in the
case of a random magnetic field \cite{KKO} as well as layered systems
in high perpendicular fields. \cite{HK}

\subsection{Universal distribution functions and its
boundary condition dependence}

The above fractal structure at the transition leads to novel
level spacing distribution $P(s)$ and conductance
distribution $P(g)$ at the critical point.
Recently, it was pointed out that $P(s)$ and $P(g)$ depend on
the boundary condition \cite{BMP,SO3}.
This was rather unexpected, since even completely different models
show the same $P(s)$ at the transition \cite{hofstetter}.

\begin{figure}
\centerline{\resizebox{7cm}{7cm}{\includegraphics{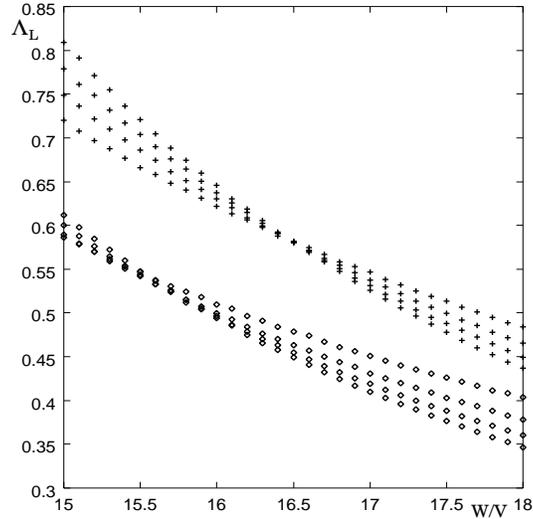}}}
 \caption{$\Lambda_L$ vs. $W$ for p.b.c. (+) and f.b.c. ($\diamond$).
The size $L$ is 6,8,10 and 12.}
 \label{lambda_eps_fig}
\end{figure}

To investigate the origin of the boundary condition dependence,
we study $\Lambda_L$ with the fixed boundary condition (f.b.c.)
imposed in the transverse direction
of a long bar, and compare it with that
in the case of periodic boundary condition (p.b.c.).
For simplicity, we concentrate here on the Anderson model with
orthogonal symmetry. 
In Fig. \ref{lambda_eps_fig}, we show the plot of
$\Lambda_L$ vs. the strength of randomness
$W$ for p.b.c. (+) and f.b.c. ($\diamond$).
The size $L$ is 6,8,10 and 12, and the accuracy of raw data is
$0.1\%$.
The critical point should be indicated by a common crossing point
in the data.
We see that this is not clear for f.b.c. but seems to occur at
$W\approx 15.5V$, while data for p.b.c. indicate $W\approx 16.5V$.
This difference is physically unacceptable.
In fact, a detailed study using $\chi$-square fitting makes clear that
the simple single parameter scaling
(\ref{eq_sps}) fails.

\begin{figure}
\centerline{\resizebox{7cm}{7cm}{\includegraphics{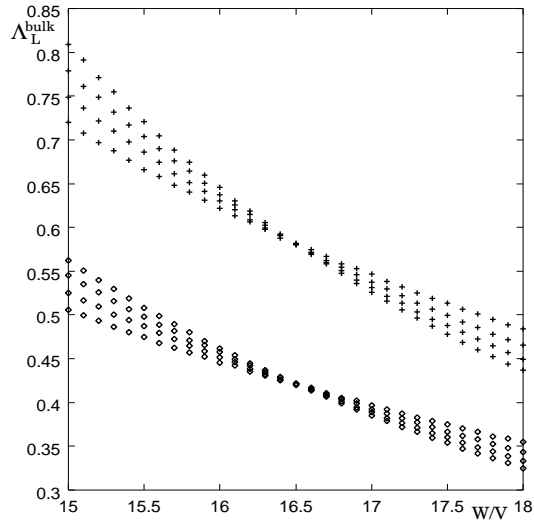}}}
 \caption{Same as Fig.\ref{lambda_eps_fig} but
surface corrections are removed.}
 \label{lambda_bulk_eps_fig}
\end{figure}

To overcome this, we have to introduce corrections to scaling 
\cite{huckestein,SO2} to take account of surface effects
\begin{equation}
\Lambda_L=f(\delta w L^{1/\nu}, b(W)L^{-1})
\label{eq_correc}
\end{equation}
with $b(W)$ an analytic function of $W$ around $W_{\rm c}$
and $\delta W=(W-W_{\rm c})/W_{\rm c}$.
We then expand this expression as
\begin{equation}
\Lambda_L=f_{\rm bulk}(\delta w L^{1/\nu})+
f_1(\delta w L^{1/\nu})b(W)L^{-1}+f_2(\delta w L^{1/\nu})
\frac{(b(W)L^{-1})^2}{2}+\cdots.
\label{eq_expansion2}
\end{equation}
Fitting the data to this expression neglecting higher order terms,
we define the bulk part of $\Lambda_L$ 
\begin{equation}
\Lambda_L^{\rm bulk}=f_{\rm bulk}(\delta w L^{1/\nu})
=\Lambda_L-
f_1(\delta w L^{1/\nu})b(W)L^{-1}-f_2(\delta w L^{1/\nu})
\frac{(b(W)L^{-1})^2}{2}.
\label{eq_bulk}
\end{equation}
Fig.\ref{lambda_bulk_eps_fig} shows the results.
Now a common crossing for f.b.c. is obtained at $W/V=16.53\pm .22$, and
the critical exponent is estimated to be $1.61\pm .16$,
consistent with the p.b.c. case.

The above results indicate that even in the limit of infinite system size,
$\Lambda_{\rm c}^{\rm bulk}$ {\it depends on the boundary condition.}
$\Lambda_{\rm c}^{\rm bulk}$ is $0.576\pm .001$ for p.b.c. \cite{SO2}
and $0.419\pm .03$ for f.b.c.
This means that , when f.b.c. is adopted, the correlation length at the
critical point for finite $L$ is significantly smaller,
leading to the fact that $P(s)$ as well as $P(g)$ tends to
share more features of an insulator.
In fact $P(s)$ with f.b.c. is closer to a Poisson distribution \cite{BMP},
and the mean conductance $\langle g \rangle$ is smaller when f.b.c.
is imposed \cite{soukoulis,SO3}.

\section{Summary and concluding remarks}
%
%
In this paper, we reviewed recent progress on the numerical study of
the Anderson transition in three dimensional (3D) systems.
Now the concept of the universality class is established numerically.
%
%

In spite of the fact that we can perform simulations for larger sizes
in two dimensions,
the estimate of the exponent is not as accurate as
in 3D except for the quantum Hall case \cite{huckestein2}.
For example, in the presence of spin-orbit scattering,
the estimates of the exponent $\nu$ in 2D are rather scattered
e.g. 2.2 \cite{SZ}, 2.5 \cite{merkt} and 2.8 \cite{fastenrath,minakuchi},
which can not be distinguished from the value of $\nu=2.35\pm .03$
obtained in the quantum Hall case.
This may be due to the fact that the corrections to scaling is larger
in 2D \cite{merkt}.
Further study is necessary to confirm the concept of universality
in 2D systems.

%
%
%

\end{document}